\documentclass{elsart}
\usepackage{amssymb}
\usepackage{epsf}
\usepackage{epsfig}

\begin{document}
\newcommand {\nc} {\newcommand}
\nc {\beq} {\begin{eqnarray}}
\nc {\eol} {\nonumber \\}
\nc {\eoln}[1] {\label {#1} \\}
\nc {\eeq} {\end{eqnarray}}
\nc {\eeqn}[1] {\label {#1} \end{eqnarray}}
\nc {\ve} [1] {\mbox{\boldmath $#1$}}
\nc {\vr} {\mbox{$\ve{\rho}$}}
\nc {\dem} {\mbox{$\frac{1}{2}$}}
\nc {\cA} {\mbox{$\mathcal{A}$}}
\nc{\CAN} {\mbox{$^{13}$C($\alpha$,n)$^{16}$O}}
\nc{\fpg} {\mbox{$^{17}$F(p,$\gamma)^{18}$Ne}}


\begin{frontmatter}
\title{Cross section predictions for hydrostatic and explosive burning}

\author[Brussels]{P. Descouvemont}
\and
\author[Basel]{T. Rauscher}
\address[Brussels]{Physique Nucl\'eaire Th\'eorique et Physique
Math\'ematique, C.P. 229, \\ 
Universit\'e Libre de Bruxelles - B 1050 Brussels - Belgium}
\address[Basel]{Departement f\"ur Physik und Astronomie,
Universit\"at Basel,
CH-4056 Basel, Switzerland}
\begin{abstract}
We review different models used for reactions involved in nuclear astrophysics. The reaction rate is
defined for resonant as well as for non-resonant processes. For low-density nuclei, we describe
the DWBA method, the potential model, the $R$-matrix method, and microscopic cluster models.
The statistical model is developed for high-level densities.
Details of calculations in the low- and high-density regimes are
illustrated with new results concerning transfer reactions and level
densities.
\end{abstract}
\end{frontmatter}

\section{Introduction}
\label{sec:intro}
Nuclear reactions are important for studying energy generation and
nucleosynthesis in astrophysical processes \cite{Cl83,RR88}. Especially in explosive
burning scenarios, large reaction networks include mainly nuclei far
from stability which have not been studied experimentally so far, and
perhaps will never be accessible in the laboratory. This poses a challenge
to both experimentalists and especially theorists, which have to predict
reaction cross sections of more or less unknown nuclei. Another
challenge is found in the comparatively low interaction energies which
pose problems for predictions as well as for measurements already close to
and at stability. Here we provide and overview of the current status of
theory and details of the remaining challenges, illustrated by several
examples of new calculations.

Theoretical models can be roughly classified in three categories \cite{De03}: \\
$(i)$ Models involving 
adjustable parameters, such as the $R$-matrix \cite{LT58} or the $K$-matrix \cite{Hu72} methods;
parameters are fitted to the available experimental data and the cross sections are extrapolated
down to astrophysical energies. These fitting procedures of course require the knowledge of
data, which are sometimes too scarce for a reliable extrapolation. \\
$(ii)$ ``Ab initio" models, where the cross sections are determined from the wave functions
of the system. The potential model \cite{BD85}, the Distorted Wave Born Approximation (DWBA)
\cite{OS91}, and microscopic models \cite{WT77,La94} are, in principle, independent of 
experimental data. More realistically, these models depend on some physical parameters,
such as a nucleus-nucleus or a nucleon-nucleon interaction which can be 
reasonably determined from experiment only. The microscopic Generator Coordinate Method (GCM)
provides a ``basic" description of a $A$-nucleon system, since the whole information is obtained
from a nucleon-nucleon interaction. Since this interaction is nearly the same
for all light nuclei, the predictive power of the GCM is important. \\
$(iii)$ Models $(i)$ and $(ii)$ can be used for low level-density nuclei only. This condition
is fulfilled in most of the reactions involving light nuclei ($A \leq 20$). However
when the level density near threshold is large (i.e. more than a few levels per MeV),
statistical models, using averaged optical transmission coefficients, are
more suitable (see Sec.\ \ref{sec:hi}).

In Sec.\ \ref{sec:rates},
we start with useful definitions commonly used in nuclear astrophysics,
such as of the reaction rates which are the actual quantities used in
reaction networks, and their relation to cross sections and the relevant
energy range.
We then discuss different approaches and reaction mechanisms for systems
with low (Sec.\ \ref{sec:low}) and high level densities (Sec.\ \ref{sec:hi}).
In nuclear astrophysics, the relevant projectile energy range often
covers both the low and high density regimes. The
difficulties concerning the prediction of reaction rates bridging the
two regimes are briefly addressed in Sec.\ \ref{sec:trans}. Concluding remarks are
presented in Sec.\ 5.

\section{Reaction Rates}
\label{sec:rates}
Let us consider a reaction between two nuclei with masses 
$A_1 m_N$ and $A_{2} m_N$ and charges $Z_{1}e$ and $Z_{2}e$  (we express here the masses in units 
of the nuclear mass $m_N$).
The theory of stellar evolution involves the reaction rate at temperature $T$, defined as \cite{FCZ75,Cl83,RR88}
\beq
N_A <\sigma v> &= & N_A \left( \frac{8}{\pi \mu m_N (k_BT)^3}\right)^{\frac{1}{2}}\ \int \sigma (E)\, E\, 
\exp(-E/k_BT)\, d E,
\label{eq_rr1}
\eeq
where we assume that the star can be considered as a perfect gas following the Maxwell-Boltzmann distribution.
In Eq. (\ref{eq_rr1}), $v$ is the relative velocity, $N_A$ is Avogadro's number, $\mu$ is 
the dimensionless reduced mass, 
and $k_B$ is Boltzmann's constant. At sub-coulomb energies the cross section varies as
\beq
\sigma(E) \sim \exp(-2\pi \eta)/E,
\label{eq_rr2}
\eeq
where $\eta=Z_1Z_2e^2/\hbar v$ is the Sommerfeld parameter.

Using (\ref{eq_rr2}), the integrand of (\ref{eq_rr1}) can be approximated
by a Gaussian shape \cite{FCZ75,Cl83} with a maximum at the Gamow peak. The energy and width are
given by 
\beq
&&E_0  =  \left[\pi \frac{e^2}{\hbar c} Z_1 Z_2 k_B T (\mu m_Nc^2/2)^{1/2}\right]^{2/3}
\approx  0.122\, \mu ^{1/3}(Z_1 Z_2 T_9)^{2/3}\  {\rm MeV},\\
&&\Delta E_0 = 4(E_0k_BT/3)^{\frac{1}{2}} 
 \approx 0.237\,(Z_1^2 Z_2^2 \mu )^{1/6}\, T_9^{5/6}\  {\rm MeV},
\label{eq_rr4}
\eeq
where $T_9$ is the temperature expressed in $10^{9}K$. The Gamow energy defines the energy range 
where the cross section needs to be known to derive the reaction rate. In most cases, this energy is much lower than the Coulomb
barrier which means that the cross sections drop to very low values.    
To compensate the fast energy dependence of the cross section, nuclear astrophysicists use the $S$-factor defined as
\beq
S(E)=\sigma(E) \, E \, \exp(2\pi\eta),
\label{eq_rr5}
\eeq
which is mainly sensitive to the nuclear contribution of the cross section. For non-resonant reactions, the energy dependence of the $S$ factor is smooth.

Rigorously the reaction rate should be calculated numerically by using
experimental or theoretical cross sections. However, the analytical approach provides a more
intuitive understanding
of the physics, and is still widely used. 
Let us start with non-resonant reactions, where the $S$-factor weakly depends on energy. In this case, the integral (\ref{eq_rr1})
can be replaced by an accurate analytical approximation. A Taylor expansion near $E_0$ provides
\beq
\exp(-2\pi\eta-E/k_BT)\approx \exp(-3E_0/k_BT)\exp\left(-\left( \frac{E-E_0}{\Delta E_0/2} \right)^2 \right).
\label{eq_rr6}
\eeq
Assuming a linear variation of $S(E)$ in the Gamow peak \cite{Cl83}, the reaction rate is then given by
\beq
N_A <\sigma v> &\approx & N_A \left( \frac{32 E_0}{3 \mu m_N (k_BT)^3}\right)^{\frac{1}{2}} \exp \left( -\frac{3E_0}{k_BT} \right)\,
 \left(1+\frac{5k_BT}{36E_0} \right)\nonumber \\
 & & \times S(E_0+\frac{5}{6}k_BT), 
\label{eq_rr7}
\eeq
which presents a fast variation with temperature.

The formalism developed here is valid for any form of the $S$-factor. The analytical procedure can be extended
further by assuming a quadratic form of the $S$-factor
\beq
S(E) \approx S_0 + S'_0E+\frac{1}{2}S''_0 E^2,
\label{eq_rr8}
\eeq
which is used in some astrophysics tables \cite{FCZ75}. We have
\beq
&&N_A <\sigma v> \approx  N_A \left( \frac{32 E_0}{3 \mu m_N (k_BT)^3}\right)^{\frac{1}{2}} \exp\left( -\frac{3E_0}{k_BT} \right) \nonumber \\
& & \times \left[ S_0 (1+\frac{5k_BT}{36E_0}) + S'_0 (E_0+\frac{35}{36}k_BT)+\frac{1}{2}S''_0 E_0 (E_0+\frac{89}{36}k_BT) \right].
\label{eq_rr10}
\eeq
This yields the well known  $T^{1/3}$ expansion of the reaction rate, up to $T^{5/3}$ \cite{FCZ75};
Eq.(\ref{eq_rr10}) provides the rate from the $S$-factor properties at zero energy.

For resonant reactions, the $S$-factor is assumed to have a Breit-Wigner form
(for the Breit-Wigner expression for cross sections see Eq.\
\ref{eq:breit}). The general definition (\ref{eq_rr1}) is of course
still valid. However, one has to account for the fast variation of $S(E)$ near the resonance energy.
Since a numerical approach is difficult for narrow resonances, we present an analytical method, widely used
in nuclear astrophysics. 

A careful analysis of integrand (\ref{eq_rr1}) shows that it always presents two maxima: at the resonance
energy, and at the Gamow energy. The peak at the resonance energy  does not depend on temperature.
The second peak, corresponding to the Gamow energy, moves according to the temperature. From these considerations, and except in the temperature range where both peaks overlap, the resonant reaction
rate can be split in two terms
\beq
N_A <\sigma v> \approx N_A <\sigma v>_R \, + \, N_A <\sigma v>_T,
\label{eq_rr11}
\eeq
where $N_A<\sigma v>_R$ corresponds to the maximum at $E=E_R$. For a narrow resonance, we have
\beq
N_A <\sigma v>_R=N_A \left( \frac{2\pi}{\mu m_N k_B T} \right) ^\frac{3}{2} \hbar^2 \omega \gamma \exp(-E_R/k_BT),
\label{eq_rr12}
\eeq
where the resonance strength $\omega\gamma$ is defined by
\beq
\omega \gamma=\frac{2J+1}{(2I_1+1)(2I_2+1)}\, \frac{\Gamma_1 \Gamma_2}{\Gamma_1 + \Gamma_2},
\label{eq_rr13}
\eeq 
$J$ being the spin of the resonance, and ($\Gamma_1,\Gamma_2$) the widths in the entrance and exit channels.
In capture reactions, the $\gamma$ width is in general much lower than the particle width.
The resonance strength is then proportional to the lower partial width.

The second maximum of integrand (\ref{eq_rr1}) yields the so-called "tail resonance" term $N_A <\sigma v>_T$.
Its analytical expression is identical to the non-resonant rate (\ref{eq_rr7}) with a Breit-Wigner
expression for $S(E)$.

\section{Systems with low level densities}
\label{sec:low}

\subsection{Introduction}
\label{sec:direct}
Although it is in general not trivial to draw a distinct line between
the direct and the compound reaction mechanism (as, e.g., in the case of
processes involving only few interaction steps of a few nucleons without
participation of the bulk of nucleons), direct reactions are
usually assumed to be fast, one-step processes with few degrees of
freedom in which the captured particle directly enters the final state
without sharing any energy with other nucleons \cite{sat83}.
It has long been known that direct reactions mainly occur at high
projectile energies where the formation of a compound nucleus is
suppressed due to the slow reaction time-scale \cite{sat83,glen83}.
Such high projectile
energies ($E_\mathrm{proj}\gtrsim
20$ MeV) are not important in the effective
energy range relevant for hydrostatic or explosive astrophysical
burning. However, direct reactions can also dominate the cross section
at very low energies in systems with low level densities. This occurs
for light nuclei and for nuclei with low particle separation energies.
For example, very neutron-rich nuclei far off stability have
decreasing neutron separation energies which implies that low level
densities are encountered in the projectile-target system at $r$-process
energies. Similar considerations hold for proton-rich nuclei close to
the proton dripline encountered in the $rp$-process. A special case concerns
nuclei with magic nucleon numbers. Due to their wide level spacing,
direct reactions may become important for intermediate and heavy nuclei
already close to or at the line of stability (see also Sec.\
\ref{sec:statmod}). On the other hand, reactions in light nuclei are
always direct because the reaction systems contain only few nucleons.
Due to this fact, reactions in solar burning 
have been known to proceed via the
direct mechanisms since decades. More recently, a number of studies have
analyzed reactions in astrophysical burning (for both capture and transfer
reactions) and also investigated
direct interactions in neutron capture on intermediate and heavy nuclei
far off stability occurring in the
$r$-process \cite{rau92,win92,kra93,mohr03,gor98}. 

In fact, the distinction between direct and compound reactions might be
misleading, especially in light nuclei. For example, it is often assumed
that the term ``direct'' is synonymous to ``non-resonant''. This is
incorrect since potential resonances can still be included in the models
described below (e.g.\ \cite{rau92}). 
Only very narrow single-particle resonances are not
reproduced in models using effective optical potentials.
Therefore we prefer to address
reactions in low level density systems and such in systems with
high level densities, and to present the relevant models for cross
section predictions in this context. Nevertheless, for simplicity we
will still occasionally use the term ``direct'' in the following
discussion. However, it is to be understood as described above.

The discussion of reactions in low level-density systems is started with
distorted wave methods for transfer and capture reactions, 
belonging to the group of potential models which
describe the dynamics of a system by solving the
Schr\"odinger equation with an effective potential. The potential is of
optical-potential type (see also Sec.\ \ref{sec:opt}) but other nucleus-nucleus
potentials are used in different models due to the nature of the model
and varying approximations of
coupling to other channels \cite{sat83}.

\subsection{The DWBA method for transfer reactions}
\label{sec:dwba}
The Distorted Wave Born
Approximation (DWBA) starts from the premise that elastic
scattering is dominant and has to be treated fully, while non-elastic
events can be treated by perturbation theory. Although DWBA is a
first-order theory, the way it is usually applied is not simply first-order.
That is because optical potentials fitted to elastic scattering data may
include higher-order effects implicitly. Therefore different potentials
are needed for higher-order methods, such as coupled-channels
calculations, than those used in DWBA
in order to reproduce the same elastic data.

The DWBA method can be applied to transfer reactions
\beq
a(=b+x)+A \rightarrow b+B\,(=A+x)
\label{eq_dwba1}
\eeq
and assumes that particle $x$ goes from the projectile $a$ to the target $A$ \cite{sat83}.
The cross section for reaction (\ref{eq_dwba1}) is obtained from the matrix elements
\beq
T^{\rm DWBA}=\int \int d \vr_{\alpha} d \vr_{\beta} \,  \chi^-_{\beta}(\vr_{\beta}) <\Psi_b \Psi_B|\Delta V|
\Psi_a \Psi_A> \chi^+_{\alpha}(\vr_{\alpha}),
\label{eq_dwba2}
\eeq
where the distorted waves $\chi^+_{\alpha}(\vr_{\alpha})$ and $\chi^-_{\beta}(\vr_{\beta})$, 
($\vr_{\alpha}$ and $\vr_{\beta}$ are the relative coordinates)
corresponding to the relative
motion in the entrance and exit channels, respectively, are generated by optical potentials $U_{\alpha}$ and
$U_{\beta}$. The residual interaction is defined in two different ways
\beq
\Delta V & = & V_{xA}+V_{bA}-U_{\alpha} \rm{\ \ (prior)} \nonumber \\
& =& V_{bx}+V_{bA}-U_{\beta} \rm{\ \  (post)},
\label{eq_dwba3}
\eeq
which correspond to "prior" and "post" definitions, respectively; they provide identical values for $T^{DWBA}$.
The main problem of the method is that the potentials are usually poorly known.
In general, a good approximation is to neglect $V_{bA}-U_{\alpha}$ or $V_{bA}-U_{\beta}$. The matrix element then contains distorted scattering wave
functions $\chi_\alpha$, $\chi_\beta$, and the radial bound state wave
functions of the transferred cluster. All of these wave functions are
usually numerically computed with optical potentials as shown in Sec.\
\ref{sec:opt}.

Since more realistic descriptions of nucleus $a(B)$ should involve other configurations than
$b+x\ (A+x)$, spectroscopic factors are introduced ($\mathcal{S}_a$ and 
$\mathcal{S}_B$). The DWBA cross section is therefore linked to 
the experimental cross section through
\beq
\sigma_{\rm exp}=\mathcal{S}_a\, \mathcal{S}_B \, \sigma_{\rm DWBA}.
\label{eq_dwba4}
\eeq

A recent work \cite{AD04} aims at investigating the precision of the DWBA 
for transfer reactions at low energies. We have 
investigated the $\CAN$ reaction by a microscopic model, and used the DWBA in conditions as close
as possible to the reference calculation.  The conclusion is twofold. On one hand, the DWBA method turns out to be very
sensitive to the conditions of the calculations: choice of the nucleus-nucleus potentials and, to
a lesser extent, of the internal wave functions of the colliding nuclei. This sensitivity is due to very
basic properties, i.e. the short-range character of the DWBA matrix elements, which are quite
sensitive to details of the wave functions. On the other hand, the difference between the DWBA
and the reference microscopic method can be fairly large, and varies with angular momentum. 
This is most likely due to antisymmetrization effects which are approximately included in the
DWBA through the choice of deep nucleus-nucleus potentials.
This
property should also occur in other systems and suggests that the DWBA method can only provide
transfer cross sections with a non-negligible uncertainty.

Future experiments with unstable beams will provide valuable
information on the level structure and spectroscopic factors (also for
the important neutron capture reactions discussed in Sec.\
\ref{sec:capture}).
Usually, (d,p) reactions are used for this purpose as neutron
capture at low energies cannot be studied on radioactive isotopes.
However, an analysis of the results might be complicated because
the experimental cross sections have to be compared to
calculations which still bear considerable uncertainties in the
optical potentials used. One has to be extremely cautious in
interpreting such comparisons as the resulting spectroscopic
factors may carry large uncertainties. A promising way out might
be the use of the adiabatic approximation instead of DWBA which
arises from neglecting the internal excitation energies of states
relative to the bombarding energy \cite{sat83,dro55}. It
can be formulated in such a manner that no optical potentials are
needed \cite{aus65}. Some parameters have to be adjusted by including elastic
data but no further reaction information is required. Further
investigation will show whether this approach is practicable for
reactions far off stability.

\subsection{Capture reactions in a potential model}
\label{sec:capture}
Capture reactions can be treated in a first-order approach involving an
electromagnetic operator describing the emission of $\gamma$-radiation
due to the dynamics in the movement of the electric charges.
Since the application to neutron or proton capture 
dominated by electric dipole
transitions is most important in astrophysics,
here we give the expression for E1 direct capture. The
cross section of the direct radiative capture process $a+A \rightarrow
\gamma+B$, which is entirely electromagnetic, is given by
\cite{RR88,Chr61,Rol73}:
\begin{eqnarray}
\label{DC}
\sigma_{\rm E1} & = & \frac{16\pi}{9}
\left(\frac{E_\gamma}{\hbar c}\right)^3
\frac{e^2\mu_\alpha^3}{\hbar^2k_\alpha}
\frac{3}{\left(2I_{\rm a}+1\right)\left(2I_{\rm A}+1\right)}
\left(\frac{Z_{\rm a}}{m_{\rm a}} -\frac{Z_{\rm A}}{m_{\rm A}}\right)^2
C^{2}\mathcal{S}_{\ell_ \beta J_\beta}\\\nonumber
&& \times \sum_{\ell_ \alpha J_\alpha}
\left(2J_\beta+1\right)\left(2J_\alpha+1\right)
\max\left(\ell_\alpha,\ell_\beta\right)\\\nonumber
&&\times \left\{\begin{array}{ccc}1 & \ell_{\beta} & \ell_{\alpha}\\
I & J_{\alpha} & J_{\beta}\end{array}\right\}^2
a_{I}^2 \left\vert R_{1 \beta \alpha} \right\vert^2
\end{eqnarray}
with the radial integral
\begin{equation}
R_{1 \beta \alpha}=\frac{1}{k_{\alpha}} \int
u_{\beta}^{*}(r)\chi_{\alpha}(r) \;r \;dr\quad .
\end{equation}
In the above expression, the energy of the emitted photon is
$E_{\gamma}$. The charge and mass of the projectile and target
nucleus are $Z_{\rm a}$, $m_{\rm a}$, $Z_{\rm A}$ and $m_{\rm A}$,
respectively. The orbital and total angular momentum quantum
numbers of the nuclei in the entrance and exit channels are
$\ell_{\alpha}$, $J_{\alpha}$, $\ell_{\beta}$ and $J_{\beta}$,
respectively. 
The spin quantum number, the orbital
and total angular momentum quantum numbers are characterized by
$S$, $L$ and $I$, respectively, with indices a, A and B
corresponding to the projectile, target and
residual nucleus, respectively. The symbol
$\left\{\matrix{\ldots}\right\}$
is the $6j$ symbol.
The radial distorted wave in the entrance channels is given by
$\chi_{\alpha}$. The radial bound state wave function of the final
state is denoted by $u_{\beta}$. The spectroscopic factor and the
isospin Clebsch--Gordan coefficient for the partition $B=A+a$ are
given by $C$ and ${\mathcal{S}}_{\ell_{\beta} J_{\beta}}$,
respectively. 

The required wave functions $\chi_\alpha$,
$u_\beta$ are found by solving the Schr\"odinger
equation with an appropriate optical potential. The spectroscopic
factor is the overlap of the wavefunctions of the target and the final
nucleus and is extracted from experiment or a more microscopic
calculation. In case of single nucleon capture the spectroscopic
factor is simply related to the occupation probability $V^2$ of
the orbital into which the nucleon is captured, e.g., ${\mathcal{S}}=
(1-V^2_j)$ for depositing a nucleon onto an even target \cite{glen83}.
The same spectroscopic factor applies to (d,p) and (d,n) reactions,
respectively. 
For a discussion of how to extract spectroscopic factors for neutron
or proton capture such transfer reactions, see the
end of the previous Sec.\ \ref{sec:dwba}.

The optical potential can be derived using the
folding approach (see Sec.\ \ref{sec:opt}) and by determining the
parameter $\lambda$ from systematics of volume integrals \cite{atz96}.
The capture cross sections are very sensitive to excitation energy
and spin of the final states. This poses a big problem far off
stability where there is no experimental information available. It
has been shown that neutron capture cross sections predicted
utilizing input from different microscopic structure models can
differ by orders of magnitude \cite{rau98}. Large jumps in the cross section
when going along an isotopic chain were also found, due to the
fact that a weakly bound state (which nevertheless dominates the
cross section) can become unbound in the subsequent isotope and
become unavailable for capture.

\subsection{R-matrix}
\label{sec:rmat}
In the $R$-matrix method, the energy dependence of the
cross sections is obtained from Coulomb functions, as expected from the Schr\"odinger equation.
The goal of the $R$-matrix method
\cite{LT58} is to parameterize some experimentally known
quantities, such as cross sections or phase shifts, with a small
number of parameters, which are then used to extrapolate the cross
section down to astrophysical energies. 

The $R$-matrix
framework assumes that the space is divided into two regions: the internal region (with radius $a$),
where the nuclear force is important, and the external region, where the interaction between the 
nuclei is governed by the Coulomb force only. 
Although the $R$-matrix parameters do depend on the channel radius $a$, the sensitivity of the
cross section with respect to its choice is quite weak.
The physics of the internal region is parameterized by a number $N$ of poles,
which are characterized by energy $E_{\lambda}$ and reduced widths
 $\gamma_{\lambda \alpha}$. In a multichannel problem, the $R$-matrix
at energy $E$ is defined as
\beq
R_{\alpha \beta}(E) = \sum_{\lambda=1}^N \frac{\gamma_{\lambda \alpha}\gamma_{\lambda \beta}}{E_{\lambda}-E},
\label{eq_rmat1}
\eeq
which must be given for each partial wave $J$. Indices $\alpha$ and $\beta$  refer to the channels.

The $R$-matrix method can be applied to transfer as well as to capture reactions.  It is usually used to investigate 
resonant reactions but is also suited to describe non-resonant processes \cite{AD98}. 
In the latter case, the non-resonant behavior is simulated by a high-energy pole, referred to as
the background contribution, which makes the $R$-matrix nearly energy independent.
The pole properties are associated with the physical energy and width of resonances, but not strictly
equal. This is known as the difference between ``formal" parameters and
``observed" parameters, deduced from experiment. In a general case, involving more
than one pole, the link between those two sets is not straightforward; recent works \cite{AD00,Br02}
have established a general formulation to deal with this problem.

\begin{figure}[t]
\centerline{\epsfig{file=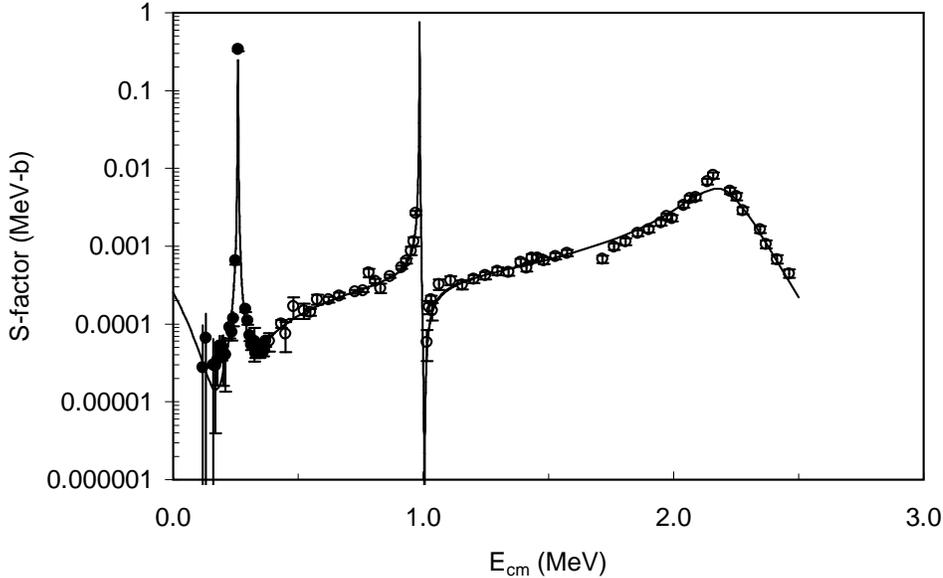,width=13cm,clip=}}
\caption{$R$-matrix fit of the $^{14}$N(p,$\gamma$)$^{15}$O reaction for 
capture to the $^{15}$O ground-state. Open circles are the data of Schr\"oder {\sl et al.} \protect\cite{SBB87} and full circles represent the LUNA data \protect\cite{FIC04}.
\label{fig_rmat}}
\end{figure}

Figure \ref{fig_rmat} shows a recent application to the $^{14}$N(p,$\gamma)^{15}$O reaction (ground-state contribution) with the data obtained by the LUNA collaboration \cite{FIC04} .  
Within the $E1$ multipolarity, $1/2^+$ and $3/2^+$ resonances are taken into account.  The 
ground-state contribution represents about 15\% of the total $S$ factor. At zero energy,
we have $S(0)=1.7\pm 0.2$ keV-b, in agreement with previous results.

\subsection{Microscopic models}
\label{sec:micro}
Microscopic models are based on basic principles of quantum mechanics, such as the treatment of all nucleons,
with exact antisymmetrization of the wave functions. The hamiltonian of an $A$-nucleon system is
\beq
H = \sum^A_{i=1} T_i \ + \ \sum^A_{i<j=1} V_{ij},
\label{eq_mic1}
\eeq
where $T_i$ is the kinetic energy and $V_{ij}$ a nucleon-nucleon interaction \cite{WT77,La94}.

The Schr\"odinger equation associated with this hamiltonian can not be solved exactly when $A>3$. The
Quantum Monte Carlo method \cite{PW01} represents a significant breakthrough in this direction, but
is currently limited to $A=10$. In addition its application to continuum states is not feasible for
the moment (it has been applied to the d$(\alpha,\gamma)^6$Li reaction \cite{NWS01}, but 
the $\alpha+d$ relative motion is described by a nucleus-nucleus potential).

In cluster models, it is assumed that the nucleons are grouped in clusters. We present here the
application to two-cluster systems. This provides a natural extension of the formalism to
scattering states. The internal wave functions of the clusters are denoted as $\phi_i^{I_i\pi_i\nu_i}(\ve{\xi_i})$, where $I_i$ and $\pi_i$ are
the spin and parity of cluster $i$, and $\ve{\xi_i}$ represents a set of its internal
coordinates. One defines the channel function as

\beq
\varphi^{JM\pi}_{\ell I} (\Omega_{\rho},\ve{\xi_1},\ve{\xi_2})=
\left[ Y_{\ell}(\Omega_{\rho})\otimes[\phi_1^{I_1\pi_1}(\ve{\xi_1})\otimes \phi_2^{I_2\pi_2}(\ve{\xi_2})]^I \right]^{JM},
\label{eq_mic1b}
\eeq
where different quantum numbers show up: the channel spin $I$, the relative angular momentum $\ell$,
the total spin $J$ and the total parity $\pi={\pi}_1{\pi}_2{(-)}^{\ell}$.

The total wave function is written as
\beq
\Psi^{JM\pi}= \sum_{\alpha \ell I} {\mathcal{A}}\, g^{J\pi}_{\alpha \ell I}(\rho) \ 
\varphi^{JM\pi}_{\alpha \ell I} (\Omega_{\rho},\ve{\xi^i_1},\ve{\xi^j_2}),
\label{eq_mic2}
\eeq
which corresponds to the Resonating Group (RGM) definition. Index $\alpha$ corresponds to different
two-cluster arrangements, and $\mathcal{A}$ is the antisymetrization operator. In most applications, the internal cluster wave functions 
are defined in the shell model. Accordingly, the nucleon-nucleon interaction must be adapted to this choice,
which leads to effective forces, such as the Volkov \cite{Vo65} or the Minnesota \cite{TLT77} interactions. The relative wave function $g(\rho)$ is to be determined from the Schr\"odinger equation.
In recent applications, this function is expanded over gaussian functions \cite{WT77,SV98}, which
corresponds to the Generator Coordinate Method (GCM). The GCM is equivalent to the RGM, but
is better adapted to numerical calculations, as it makes uses of projected Slater Determinants.

The main advantage of cluster models with respect to other microscopic theories is its ability to deal
with reactions, as well as with nuclear spectroscopy. The first applications were done for reactions involving
light nuclei, such as d, $^3$He or $\alpha$ particles. More recently, much work has been devoted to the improvement of
the internal wave functions: multicluster descriptions \cite{DB94,DD96}, large-basis shell model extensions
\cite{De96}, or monopolar distortion \cite{BK92}. Microscopic cluster models have a wide range of applications,
both in low-energy reactions and in the spectroscopy of light nuclei. The main limitation arises from the number of 
channels included in the wave function, which reduces the validity of the model at low energies.
Also high-level densities require many channels in the wave functions. Nuclear astrophysics is probably one of the
best candidates for applications of microscopic cluster models. The low energies and the low densities involved in most 
nuclei make the conditions of application quite valid. 

We illustrate cluster models with a recent application to the $\fpg$ reaction \cite{DD04}.
The knowledge of the \fpg\ reaction rate  is important for the understanding
of the energy production and of the nucleosynthesis in novae \cite{I02}.
Until now,  a direct measurement of the cross section down to astrophysical 
energies (typically 0.3 MeV) has not been performed. 
It is now well established that the $3^+_1$ level
dominates the rate  at high temperatures since
it is the only  $s-$wave resonance at low energies. 
The gamma width remains experimentally unknown and is
estimated  by shell-model calculations.
At low temperatures, the non-resonant contribution is dominant 
but is evaluated with simple models \cite{Gar91} which need confirmation from
more elaborate theories.  
\begin{figure}[t]
\centerline{\epsfig{file=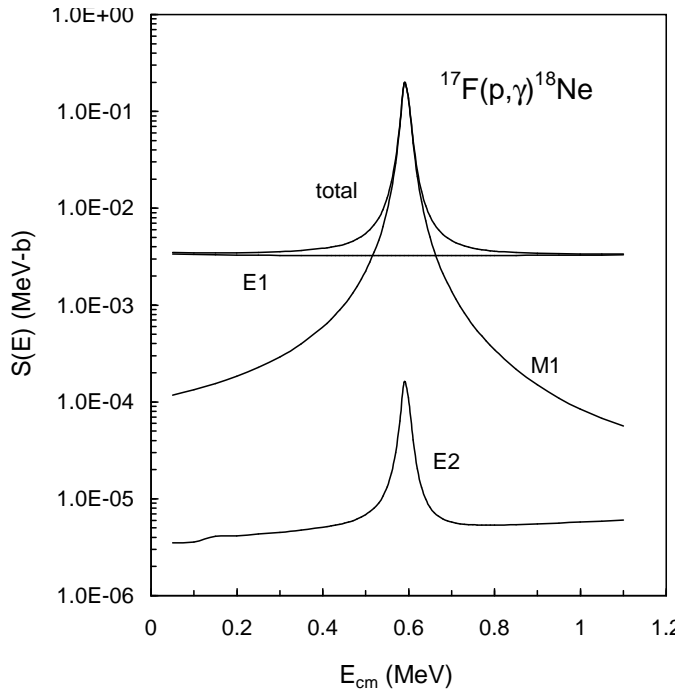,width=9cm,clip=}}
\caption{$\fpg$ astrophysical $S$-factor with the contribution of different multipolarities.
\label{fig:fpg}}
\end{figure}

The GCM wave functions are defined as
a combination of $^{17}$F + p and $\alpha + ^{14}$O channel functions.
The $^{17}$F and the $^{14}$O nuclei  are described by
$s, p$ and $sd$ (for $^{17}$F) harmonic oscillator shells.
This framework allows the  description   of the
$^{17}$F$(5/2^+,1/2^+,3/2^+)$+p and $^{14}$O$(0^+,2^+) + \alpha$ channels. 

The proton width of the $3^+$ resonance (21 keV) is in fair agreement  with the experimental value ($18\pm 3$ keV) \cite{Bard00}. For the gamma width, we find 33 meV;   
a recent evaluation of the reaction rate \cite{Bard00} uses $\Gamma_{\gamma}=29 \pm 19$ meV.
The astrophysical $S$-factor  is  calculated for the $E1$,
$E2$, and $M1$ multipolarities. In each case, the four  
$0^+_1$,$2^+_1$, $2^+_2$ and $4^+_1$ bound states are  considered. 
The $J$ values for scattering states are  limited to 4, which corresponds to $\ell=2$.
The total and partial  $S$-factors are displayed in Fig. \ref{fig:fpg}.

\section{Systems with high level densities}
\label{sec:hi}

\subsection{Optical model}
\label{sec:opt}
Due to the complexity of the nucleon-nucleon (NN) interaction, one often
resorts to working with effective interactions instead of
solving microscopic models based on NN potentials (but see Sec.\
\ref{sec:micro}). Widely used in calculating different reaction
mechanisms is the optical model. In that model, the complicated
many-body problem posed by the interaction of two nuclei is replaced by
the much simpler problem of two particles interacting through an
effective potential, the so-called optical potential
\cite{sat83,glen83,gh92}.
Such an approach is usually feasible only with few
contributing channels. Always included is
elastic scattering. That is why optical potentials can be derived from
elastic scattering data.

In general, the optical potential is complex. The real part describes
the refraction of the incident waves by the nucleus (elastic
scattering), the imaginary part their absorption, hence the term
``optical'' model. Thus, the imaginary
part of the resulting wave function contains all non-elastic processes
removing flux from the incident channel. In general, the potential is
expected to be energy-dependent for obvious reasons. Most applications
make use of local potentials although, in principle, also the effective
potential should be non-local. An additional slight energy-dependence is
introduced by this neglection \cite{sat83}. 

The required potential parameters
can be derived either by fits to experimental elastic
data or from microscopic considerations. An example of the latter is the
folding approach \cite{sl77} which is often employed to determine the real
potential. In this approach an effective NN interaction is folded with
projectile and target nucleon density distributions. The latter have to
be derived from electron scattering data or further microscopic considerations.
Another example for a more microscopic approach is the use of the local
density approximation \cite{jlm77,bau01}. The advantage of this model
is that also the imaginary part can be determined. Again, certain
parameters have to be fitted to experimental data and density
distributions have to be known.

The standard definition of the optical model stems from the description
of elastic and inelastic scattering.
The time-independent radial Schr\"odinger equation is numerically solved
with an optical potential which provides a mean interaction potential,
averaging over individual NN interactions. In consequence,
single-particle resonances cannot be described in such a model. However,
resonances stemming from potential scattering can still be found.
Elementary scattering theory yields expressions for the elastic cross
section and the reaction cross section. The latter includes all
reactions and inelastic processes which cause loss of flux from the
elastic channel. With the diagonal element $S^{\alpha \alpha}$ of the
S-matrix $S$ (sometimes also called scattering
matrix or collision matrix),
the reaction cross section for spinless particles
is then given by \cite{gh92}
\begin{equation}
\label{eq:csreac}
\sigma^{\alpha \alpha}_\mathrm{r} = \frac{\pi}{k^2}
\sum_L \left(2L+1\right) \left( 1-\left| S^{\alpha \alpha}_L\right|^2
\right) \quad .
\end{equation}
This can be generalized to other channels $\beta$, not just the elastic one.
The elements of the S-matrix are complex, in general, and related to the
scattering amplitude $f$ of the outgoing wave function
\begin{equation}
f^{\alpha \beta}=\frac{1}{2ik}\sum_L (2L+1) (S^{\alpha \beta}_L
-1)P_L(\cos \theta) \quad ,
\end{equation}
which is nothing else than the transition amplitude
$T^{\alpha \beta}=-(2\pi \hbar)/\mu_\beta f^{\alpha \beta}$, in turn,
with $\mu$ being the reduced mass. As mentioned above, 
the imaginary part of the
optical potential gives rise to an absorption term in the solution of
the Schr\"odinger equation, thus removing flux from the considered
channels. Therefore, the matrix element $S^{\alpha \alpha}$ is also
related to the transmission coefficient
\begin{equation}
t_L = \left( 1-\left| \mathrm{e}^{2i\delta_L}\right|^2 \right)
=1-\left| S^{\alpha \alpha}_L\right|^2
\end{equation}
which describes the absorption of the projectile by the nucleus.
Important for practical application is that the phase shifts $\delta_L$
can be derived from elastic data.

In the preceding Sec.\ \ref{sec:direct} we have already discussed
several models making use of potentials of the optical type.
In the following, the statistical model of nuclear reactions is
described which also applies optical potentials to calculate the
formation and decay of the compound nucleus.

\subsection{Statistical model}
\label{sec:statmod}
It is well established that the number of states in a nucleus
increases in principle exponentially with excitation energy. Thus,
with increasing projectile energy, different resonance regimes are
covered. Starting without or only few resonances at low energy and
more, but still well separated, resonances at intermediate
energies, one will eventually encounter a region where the average
resonance width $\left< \Gamma \right>$ becomes larger than the
average level spacing $D=1/\rho$. This is the domain of the
compound nucleus reaction which can be calculated in a statistical
model only using average resonance properties, also called the
Hauser-Feshbach approach. In astrophysical applications, the
compound reaction mechanism is especially important for reactions
with charged projectiles as the relevant energy range (given by
the Gamow peak, eq.\ \ref{eq_rr4}) is shifted to higher energies as compared to
neutron-induced reactions. Nevertheless, also for reactions
involving neutrons the statistical model can be applied when the
reaction $Q$-value or the level density is sufficiently high. This
is true for most intermediate and heavy nuclei close to stability
with non-magic neutron numbers. The relevant quantity is the
number of available levels in the Gamow peak. An estimate of the
applicability range of the statistical model is given in \cite{rtk97}.

The statistical model is based on the Bohr independence
hypothesis \cite{boh36}. It states that the projectile forms a compound system
with the target and shares its energy among all of the nucleons.
This implicitly requires long reaction time-scales as the compound
nucleus has to live long enough to establish complete statistical
equilibrium. Compared to the direct mechanism (Sec.\
\ref{sec:direct}) it is about 5 orders of magnitude longer and
includes many degrees of freedom. Finally, the compound nucleus
can decay by emitting photons or particles. According to the Bohr
hypothesis the final decay is independent of the formation
process, except for the basic conservation laws of energy, angular
momentum, parity, and nucleon number. Thus, the cross section can
be factorized into two terms
\begin{equation}
\sigma^{\alpha \beta} _\mathrm{CN}=\sigma^\alpha_\mathrm{form}b_\mathrm{dec}
=\sigma^\alpha_\mathrm{form}\frac{\left< \Gamma^\beta 
\right>}{\left< \Gamma_\mathrm{tot}\right>}\quad .
\end{equation}
the formation cross section $\sigma_\mathrm{form}$ and a
branching ratio describing the probability for decay to the
observed channel $\beta$.
An early implementation of this was the Weisskopf-Ewing theory \cite{wei40}.
Today, the Hauser-Feshbach approach \cite{hf52} is widely used which also
incorporates angular momentum conservation. Modern theories also
treat pre-equilibrium emission of particles in different ways.
However, for the energies encountered in astrophysical scenarios,
this additional mechanism is usually negligible.

In the Hauser-Feshbach theory, the formation cross section is
calculated as in the optical model (Sec.\ \ref{sec:opt})
but using averaged transmission coefficients $\left< T \right>$
which can again be computed from a Schr\"odinger equation and using
appropriate optical potentials.
The average widths occurring in the branching ratio $b_\mathrm{dec}$
are related to average transmission coefficients via
$\left< T\right>=2\pi \rho \left< \Gamma \right>$,
with $\rho$ being the average level density.
In fact, it can be
shown \cite{ctt05} that the expression for the compound reaction cross section
\begin{equation}
\label{eq:avbreit}
\sigma_\mathrm{CN}^{\alpha \beta}=\frac{\pi}{k_j^2}
\frac{1+\delta_{ij}}{(2I_i+1)(2I_j+1)}\sum_{J\pi j\ell j'\ell'}
(2J+1)\frac{\left< T^\alpha_{Jlj}\right>\left< T^\beta_{
Jl'j'}\right>}{\sum_{\alpha lj} \left< T^\alpha_{Jlj}\right>} 
W^{\alpha \beta}
\end{equation}
can be derived from a Breit-Wigner expression
and substituting the resonance parameters with average
quantities. The Breit-Wigner expression for the cross section as the sum
of $n$ individual resonances is given by
\begin{equation}
\label{eq:breit}
\sigma^{\alpha \beta}_{\rm BW}=\frac{\pi^2}{k^2_j}
\frac{1+\delta_{ij}}{(2I_i+1)(I_j+1)}\sum_{k=1}^n (2J_k+1)
\frac{\Gamma_k^\alpha
\Gamma_k^\beta}{(E-E_k)^2+(\Gamma^{\rm tot}_k/2)^2} \quad .
\end{equation}
The indices $i$ refer to the target, $j$ to the projectile, $\alpha$ to
the entrance channel, and $\beta$ to
the exit channel. The total
width of a resonant state $k$, $\Gamma^{\rm tot}_k$, is the sum over the
partial
widths of the individual decay channels
$\Gamma^{\rm tot}_k =\Gamma_k^\alpha+\Gamma_k^\beta+\dots$ Using the
relation
\begin{equation}
\left<\frac{\Gamma_k^\alpha \Gamma_k^\beta }{(E-E_k)^2+(\Gamma^{\rm
tot}_k/2)^2}\right>
=\frac{1}{\Delta E}\int\frac{\Gamma_k^\alpha \Gamma_k^\beta}{(E-E_k)^2+
(\Gamma^{\rm tot}_k/2)^2}dE
\approx \frac{2\pi}{\Delta E}\frac{\Gamma_k^\alpha \Gamma_k^\beta}{\Gamma^{\rm
tot}_k}
\end{equation}
we can write
\begin{eqnarray}
&\left<\sum_k(2J_k+1)\frac{\Gamma_k^\alpha \Gamma_k^\beta }{(E-E_k)^2+
(\Gamma^{\rm tot}_k/2)^2}
\right>
=\sum_{J,\pi}(2J+1)2\pi\frac{\Delta n(J,\pi)}{\Delta
E}\left<\frac{\Gamma_{J,\pi}^\alpha \Gamma_{J,\pi}^\beta}{\Gamma^{\rm
tot}_{J,\pi}}\right>
&\nonumber\\
=&\sum_{J,\pi}(2J+1)2\pi\rho_{J,\pi}
\frac{\left<\Gamma_{J,\pi}^\alpha \right>\left<\Gamma_{J,\pi}^\beta
\right>}{\left<\Gamma^{\rm tot}_{J,\pi}\right>}W^{\alpha \beta}(J,\pi)&\quad .
\end{eqnarray}
From the last line above
it is easily seen that the Hauser-Feshbach
cross section in Eq.\ \ref{eq:avbreit} is an averaged Breit-Wigner cross
section $\sigma_\mathrm{CN}=\left< \sigma_\mathrm{BW}\right>$, when
$W^{\alpha \beta}=1$.

In the general case, the width fluctuation coefficients
\begin{equation}
W^{\alpha \beta}(E,J,\pi)=\left<\frac{\Gamma^\alpha_{J,\pi}(E)\Gamma^\beta
_{J,\pi}(E)}{\Gamma^{\rm tot}_{J,\pi}(E)}\right>
\frac{\left<\Gamma^{\rm tot}_{J,\pi}(E)\right>}{
\left<\Gamma^\alpha_{J,\pi}(E)\right>\left<\Gamma^\beta_{J,\pi}(E)\right>}
\end{equation}
describe non-statistical correlations between the widths in the channels
$\alpha$ and $\beta$.
They differ from unity close to channel openings and enhance the weaker
channel \cite{gh92}.

Although it might seem tempting to conclude that the cross section
of a reaction proceeding through the compound mechanism should be
smooth because it is formed from the superposition of amplitudes
from a very large number of states with random phases, this is a
wrong assumption \cite{gh92}. Ericson \cite{eri60}
first showed that the cross sections can
continue to show large fluctuations. The usual Hauser-Feshbach
equations do not account for these fluctuations. Therefore, a
meaningful comparison to experimental data is only possible after
averaging the data over a sufficiently wide energy range,
comparable to the average resonance widths. When using the
statistical model to compute astrophysical reaction rates (or when
deriving rates experimentally directly) this is taken care of
automatically. However, when using beams with a very narrow energy
spread it should be noted that the results cannot be directly
compared to calculations \cite{koe04}.

There is a number of available predictions, e.g.\
\cite{nonsmoker,most,rath,cow91,sar82,woo78,holm76}. The
underlying model is the same in all cases \cite{hf52} but the used nuclear
properties and prediction of properties are different. Due to the
number of reactions to be included in reaction networks, global
models have to be applied. It should be noted that rates derived
from the statistical model are not applicable at all astrophysical
temperatures. The specific temperature limits are given, e.g.\ in
\cite{rath}, and it is advisable to refrain to use those rates below or
close to the given temperature.

\begin{figure}[t]
\centerline{\epsfig{file=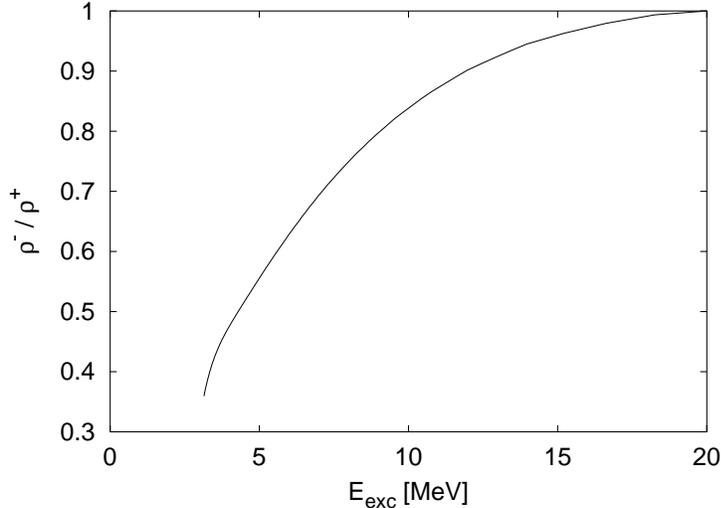,width=7cm,clip=,angle=-90}}
\caption{Shown is the nuclear level density ratio of $\pi=-1$ levels to $\pi=+1$
levels in $^{61}$Fe as a function of excitation energy.
Note that the ratio is considerably lower than 1
even several MeV above the neutron separation energy
($E_\mathrm{sep}=5.58$ MeV).\label{fig:levpar}}
\end{figure}

The most relevant quantities for calculating the statistical model
cross sections are reaction $Q$ values, optical potentials, level
densities, the electromagnetic transition strength (mostly E1 and
M1) for photon transmission coefficients, and information on
low-lying discrete states. Due to the fact that average properties
of usually a large number of transitions are used, the results are
not as sensitive to a variation in the input as the direct
reaction cross sections. Nevertheless, extrapolations to far from
stability can bear considerable errors due to effects which are not
pronounced close to stability. Modern investigations
focus on improving the prediction of nuclear properties needed in
the statistical model. There are basically two types of
approaches. First, one tries to obtain new data as close as
possible to the astrophysically relevant energy range for a number
of nuclei. From a systematic comparison of predictions with
various nuclear inputs to the data, theory is tested close to
stability. Another approach is to implement recent microscopic
information to derive nuclear properties far off stability and study
their impact on the rates. The latter approach is the only one possible
when dealing with highly unstable target nuclei but it is prone to the
highest uncertainties. An attempt has to be made not just to take
nuclear structure predictions at face value but rather to learn from
comparison of different model and to extract dependences of the
required nuclear properties. When such basic dependences are known,
they can be easily implemented, e.g., in a parametrized form, in global
reaction models. A good example for proceeding this way is the treatment
of the nuclear level density. A number of recent microscopic
calculations have proved that the backshifted Fermi-gas approach \cite{rtk97}
perfectly reproduces the shape of the level density as a function of
energy, both for exotic nuclei and to high excitation energies 
\cite{dean,paar,alha,VanI,alha1}. Thus,
the determination of the level density parameter and the backshift
remains. Those can again be extracted in a simple manner from globally
available calculations of nuclear masses. At low excitation energies the
Fermi-gas formula has to be replaced by a constant temperature formula
or a similar approach. At intermediate excitation one
has to account for the fact that parities are not distributed evenly.
Based on Monte Carlo shell model calculations a simple model for the
partition function ratio was suggested recently. We have adopted this
model and are currently extending it to all nuclei \cite{rau03,moc04}. Fig.\
\ref{fig:levpar} shows an example for the calculated parity ratio in
$^{61}$Fe.
This will allow not only to establish a better description of cross
sections for nuclei with low separation energies but will also be
important for checking the applicability range of the statistical model.

Comparison with recent reaction data has shown that the global
statistical model approaches work well at stability for reactions
involving neutrons and protons as projectiles but that there can be
problems predicting $\alpha$ capture due to the uncertainty in the
global optical $\alpha$ potential 
\cite{rtk97,bao00,ful96,sau97,som98,bork98,chlo99,haris01,gyurky01,ozkan02,galanopulos03,gyuri03}. 
The problem lies in the fact that astrophysically relevant
$\alpha$-energies are close to or below the Coulomb barrier and the
potentials show a
strong energy-dependence over this energy range. Thus, the optical
potentials
are both hard to measure and difficult to calculate.
A number of attempts was made
to derive an improved global description of the potential \cite{koe04,gal03,dem01,fuel01,gle00,moh00,rau98proc,moh97,atz96}. Very
promising seems the new global potential of \cite{avri} which led to good
agreement for a number of reactions \cite{dani}.

A new group of
experiments is testing the ground-state photon transmissions and thereby
the reliability of the used GDR strength 
\cite{mohrlett,babilon01,vogt01,vogt02,mohrgam03,utso03,vogt03,utsoprc03,sonn03,rathgam}
(see also Mohr et al., this volume).
Also there, no discrepancies
have been found so far. However, concerning the photon
strength function it has been suggested that there could be some
additional strength at low photon energy, created by soft-mode
vibrations, often called pygmy resonance. A low-lying pygmy resonance
could increase the neutron capture cross section far off stability
considerably \cite{gor98,gork}. However, different nuclear structure models 
still disagree about
position, strength and existence of this resonance in exotic nuclei,
e.g.\ \cite{Vanisa,vrete}. It will only be
important if there is additional strength below the neutron separation
energy, if the statistical model is applicable, and if it also appears
in nuclei which are sufficiently close to stability so that they are not
produced in a reaction equilibrium but rather by individual capture or
photodisintegration reactions.

\subsection{Transition regions with intermediate level densities}
\label{sec:trans}
Because of the Coulomb barrier, the Gamow window for charged particle
reactions is located at higher energy than for neutron-induced
reactions (see Sec.\ \ref{sec:rates}). Therefore the statistical model
of nuclear reactions is applicable for the majority of cases with
charged projectiles. The only exceptions are proton captures close to
the proton dripline as encountered in the $rp$ process. The situation is
different for neutron capture. Although a large number of reactions
at stability proceeds at high level densities, already at stability
there are some which reach into the regime of intermediate and low level
densities. As stated above, this happens for nuclei with low level
densities at neutron shell closures. On the neutron-rich side, unstable
nuclei exhibit low neutron separation energies soon, leading to
formation of the compound nucleus at low level density.

The effective neutron energies are quite low in all astrophysical scenarios,
$E_\mathrm{n}\lesssim 120$ keV \cite{bao00}. 
Although some scenarios start out from
higher plasma temperatures leading to higher neutron energies, reaction
equilibria are established and individual rates lose importance at such
conditions. Since one has to cover the energy range of
$E_\mathrm{n}\lesssim 120$ keV, it becomes obvious that several reaction
mechanisms might contribute in different parts of that range in nuclei
with low level density. Especially
the treatment of isolated resonances is very difficult because the
locations and strengths have to be predicted very accurately in order to
apply R-matrix or Breit-Wigner formulas. Furthermore, the interplay
between competing reaction mechanisms has not been treated rigorously so
far. It can be shown that there should be no coherent effects between
direct and compound reactions \cite{sat83} but there can be interferences
between single resonances and a direct background \cite{raurai}.

The imaginary parts of the optical potentials used might have to
be modified to account for the flux into other reaction channels.
The statistical model uses the optical model reaction cross section
as given by Eq.\ (\ref{eq:csreac}) to compute the formation cross section
$\sigma_\mathrm{form}$. Thus, it is implicitly assumed that all flux
lost from the elastic channel is used to form a compound nucleus
subsequently decaying statistically. In the transition region this is
an oversimplification as other reaction mechanisms become important,
too. Therefore, an increasing fraction of the lost flux is accounted for
in other reaction channels, not in the statistical one, and the
imaginary part used for $\sigma_\mathrm{form}$ would have to be decreased
accordingly with decreasing level density. A neglection of this effect
leads to an overestimate of $\sigma_\mathrm{form}$ and thus of the
statistical model cross section.

Although there have been successful calculations incorporating all
reaction mechanisms \cite{krausmann,v45,oak2} and explaining
differences between resonance and activation measurements 
\cite{georev,oak1} without
modification of the potential, this conceivable complication should be
kept in mind. In consequence, it is very important to have measurements
directly in the relevant energy range. Extrapolations upwards from lower
energies or downwards from higher energies can prove treacherous,
especially when different reaction mechanisms are involved
\cite{raugub}. Even when just measuring elastic data to derive optical
potentials, the energy range relevant for astrophysics has to be
covered.

\section{Conclusion}

The methods to predict nuclear reaction rates are well established.
Approaches based on nucleon-nucleon interactions have
the advantage that they are first-principle methods,
automatically including all relevant effects such
as antisymmetrization, correlations, Pauli blocking, etc. 
On the other hand, depending on the model, effective
NN interactions might have to be adjusted for each case. Common to all
microscopic approaches is that the involved numerics limits the
application to few-nucleon systems. The other class of reaction
theories, based on optical potentials derived either from microscopic
structure considerations or from phenomenology, average over certain
properties and can be viewed as approximations to the exact calculation.
Their strength lies in the wide applicability and the fact that the
neglection of certain effects can be compensated by slightly adjusted
optical potentials (which is automatically ascertained when deriving
phenomenological potentials from experiment).

However, any model can only be as good as the underlying nuclear structure
model required to provide the necessary input for the reaction
calculation. While the combination of experimental data and predicted nuclear
structure leads to good accuracy in the theoretical rates at and close
to stability, larger errors should be expected further off stability.
Future radioactive ion beam facilities (RIA, GSI upgrade) will provide
the possibility for improvement. The situation is also eased by the fact
that nuclei far off stability are usually produced in reaction
equilibria (such as the (n,$\gamma$)--($\gamma$,n) equilibrium of the
$r$ process) where individual reactions are not important. Thus,
reactions have to be explicitly included only closer to stability (hydrostatic
and explosive burning, freeze-out from high-temperature scenarios). Due
to this fact it is actually possible to produce meaningful astrophysical
models. We can also expect the future facilities to directly study this
relevant range of nuclei.

\begin{ack}
A part of this text presents research results of the Belgian program P5/07 on
interuniversity attraction poles initiated by the Belgian-state
Federal Services for Scientific, Technical and Cultural Affairs.
This work was partially supported by the Swiss National Science
Foundation (grant 2000-061031.02). T. R. acknowledges a PROFIL
professorship from the Swiss NSF (2024-067428.01).
\end{ack}

\end{document}